\documentclass{ws-ijmpc}

\usepackage{url}
\usepackage[%
  colorlinks=true,
  urlcolor=blue,
  linkcolor=blue,
  citecolor=blue
]{hyperref}
\hypersetup{linkcolor=red, citecolor=red, colorlinks=true}

\usepackage{mathtools,amsmath}
\usepackage{amsmath,amssymb}

\begin{document}

\markboth{F. L. Forgerini et. al.}
{Directed propaganda in the majority-rule model}

\catchline{}{}{}{}{}

\title{Directed propaganda in the majority-rule model}

\author{Fabricio L. Forgerini$^1$,
        Nuno Crokidakis$^2$ \and M\'arcio A. V. Carvalho$^3$
}

\address{
$^{1}$Centro de Forma\c{c}\~ao em Tecno-Ci\^encias e Inova\c{c}\~ao \\
 Universidade Federal do Sul da Bahia, \hspace{1mm} Itabuna/BA,  \hspace{1mm} Brazil\\
 fabricio.forgerini@ufsb.edu.br\\
$^{2}$Instituto de F\'{\i}sica, \hspace{1mm} Universidade Federal Fluminense \hspace{1mm} Niter\'oi/RJ, \hspace{1mm} Brazil \\
 nunocrokidakis@id.uff.br\\
 $^{3}$Centro de Forma\c{c}\~ao em Pol\'iticas P\'ublicas e Tecnologias Sociais \\
 Universidade Federal do Sul da Bahia, \hspace{1mm} Itabuna/BA, \hspace{1mm} Brazil\\
 carvalhomav@ufsb.edu.br}

\maketitle

\begin{abstract}

Advertisement and propaganda have changed continuously in the past decades, mainly due to the people's interactions at online platforms and social networks, and operate nowadays reaching a highly specific online audience instead targeting the masses. The impacts of this new media effect, oriented directly for a specific audience, is investigated on this study, in which we focus on the opinion evolution of agents in the majority-rule model, considering the presence of directed propaganda. We introduce $p$ as the probability of a ``positive'' external propaganda and $q$ as the probability to the agents follow the external propaganda. Our results show that the usual majority-rule model stationary state is reached, with a full consensus, only for two cases, namely when the external propaganda is absent or when the media favors only one of the two opinions. However, even for a small influence of external propaganda, the final state is reached with a majority opinion dominating the population. For the case in which the propaganda influence is strong enough among the agents, we show that the consensus can not be reached at all, and we observe the polarization of opinions. In addition, we show through analytical and numerical results that the system undergoes an order-disorder phase transition that occurs at $q_c = 1/3$ for the case $p = 0.5$.

\end{abstract}

\section{\label{sec:intro}Introduction}

\qquad The challenge of understanding the collective behavior emerging from individuals in society has fascinated physicists for the last few decades and their contributions gave rise to what have been called sociophysics, an attempt to explain how a social collective behavior emerges from individual agents \cite{Schweitzer,nuno_csf,galam}. However, the lineage of the agent-based models can be traced back to the 1940’s, with the theoretical conception of Von Neumann’s self-replicating machines \cite{von_neumann} and Sakoda’s general discrete dynamical model for social interactions \cite{sakoda}. It is important to note that both developments emerged alongside the early computers, and, on several occasions, the initial simulations were conducted without their assistance.

Von Neumann’s devices would follow instructions in order to create identical copies of themselves. Stanislaw Ulam, a mathematician and close associate of von Neumann, further refined this idea by suggesting that the machine could be represented on paper as a grid of cells, giving rise to what would later be termed cellular automata. Intrigued by Ulam's proposition, von Neumann proceeded to develop the initial design, thus pioneering the concept of cellular automata.

The foundational Sakoda model encompasses the dynamics of social interactions within a network involving two distinct groups of individuals. The model incorporates social specific attitudes such as attraction, repulsion, and neutrality, which guide the evolution of social bondings. Each individual in the network evaluates its social expectations across all feasible locations, displaying a preference for proximity to individuals associated with positive attitudes of attraction while avoiding locations near those linked to negative attitudes of repulsion. This evaluative process is randomly iterated across all individuals, leading to the recursive implementation of Sakoda's algorithm. When certain conditions are met, this iterative process drives the system towards the emergence of a well-organized spatial pattern.

Departing from the paradigm of von Neumann's machine, John Conway's Game of Life \cite{Gardner} functioned based on elementary rules operating within a simulated realm portrayed as a two-dimensional field. Furthermore, the emergence of the agent-based model as a framework for studying social systems can be largely attributed to the efforts of computer scientist Craig Reynolds \cite{Reynolds}, who  sought to model the dynamics of living biological agents, commonly referred to as artificial life. But it was the groundbreaking model of Sugarscape, developed  by Joshua M. Epstein and Robert Axtell \cite{Epstein}, which served as a platform for conducting large-scale simulations and investigations. It enabled the exploration of intricate dynamics encompassing a wide range of social phenomena, such as seasonal migrations, pollution, sexual reproduction, combat, disease transmission, and cultural diffusion.

In 1999, Gilbert Nigel, in collaboration with Klaus G. Troitzsch, made a significant scholarly contribution by publishing the inaugural textbook on social simulation, entitled ``Simulation for the Social Scientist" \cite{Nigel}. This pioneering work provided a comprehensive foundation for understanding and applying simulation techniques in the context of social science research.

Kathleen M. Carley made a crucial contribution in promoting the utilization of simulation methodologies in the context of organizational systems. Her scholarly work, ``Computational Organizational Science and Organizational Engineering" \cite{Carley}. Following these initial advancements, a proliferation of models, methodologies, and scholarly publications in the field of social simulation ensued. The topics have also diversified, encompassing knowledge transmission, institutions, reputation, social norms, elections, economics, among many others. Of particular relevance to this study is the theme of advertisement and propaganda, or belief transmission. A variety of sociological relevant problems have been studied by the means of the statistical mechanics techniques jointly with the massive use of computational modeling and simulations \cite{Castellano}. Different realms from social sciences such as opinion \cite{Redner,Sznajd,galam1,Crokidakis,pmco_jstat} and language dynamics \cite{Schwämmle,BARONCHELLI}, community formation and detection \cite{Newman1,Newman2}, gossip and rumor spreading \cite{Haeupler,Herrmann,Huo}, to name a few, have been extensively studied by physicists in collaboration with scientist from different areas. 

The basic starting point of thermodynamics and statistical mechanics focus on the study of an isolated system with interacting particles, in which the macrostates of the system are described by a set of parameters of its microscopic constituents. By direct analogy from the statistical physics one can apply the methods used in the framework of the statistical mechanics to the study of social problems, such as opinion dynamics, in which the individuals can interact with other individuals as the particles do in the standard thermodynamics systems \cite{Estrada_Knight,durlauf}.

The traditional advertisement and propaganda have enormously changed in the last decades, from the usual mass media that reach a broad audience with same type of information, to the directed advertisement, mediated by complex algorithms and artificial intelligence operating in social media and online platforms on the web \cite{Chomsky,Nuno}. This new form of how the information is reaching people is poorly studied in the context of opinion dynamics and yet it is not well understood how it can affect the way consensus (or majority) is formed in a population \cite{Gimenez,Rodriguez,galam_new}.

Recently, researchers of several areas are showing interest in models of opinion diffusion in social media and polarization formation. Examining the dissemination of news topics on Twitter, the authors in \cite{arcila2019modeling} discusses through a topic modeling algorithm that the data suggests that there is a relationship between the most-mentioned profiles, the more frequent keywords and the main underlying news topics. In Ref. \cite{helfmann2023influencers}, the authors studied how strategies of influencers to gain more followers can influence the overall opinion distribution. They found that moving towards extreme positions can be a beneficial strategy for influencers to gain followers in social medias. Yet regarding information dissemination, the authors in \cite{burbach2020opinion} studied which factors influence whether a user of online social networks disseminates information or not. They found that the network type has only a weak influence on the distribution of content, whereas the message type has a clear influence on how many users receive a message. A recent work investigated how main stream media signed interaction might shape the opinion space. The authors focused on how different size (in the number of media) and interaction patterns of the information system may affect collective debates and thus the opinions' distribution. The results show that plurality and competition within information sources lead to stable configurations where several and distant cultures coexist \cite{quattrociocchi2014opinion}. Another recent work studied the dynamics of opinion diffusion in social networks, where the authors considered post transmission and post distribution, representing the users' behavior and the social network algorithm, respectively. The results that the dynamics can converge to consensus formation or to polarization. It was also found that friendship rewiring helps promote echo chamber formation, and that the social network algorithm is crucial to mitigate or promote polarization \cite{FERRAZDEARRUDA2022265}. For recent reviews of other similar models, see \cite{RAZAQUE20221275,caled2022digital}.

We are interested here in the majority-rule model \cite{galam1,GALAM1999132,PhysRevLett.90.238701}. It was originally proposed by Serge Galam to describe how a fully-connected population composed by $N$ individuals reach consensus, in which the agents can decide in favor or against a particular subject \cite{Galam2}. A random number of agents, called ``discussion group", is selected to debate, interacting among each other and, after the discussion, all of them follow the most popular opinion, the majority opinion of the group. The results for the original model show that the initial majority opinion win the debate after some time, i.e., the steady states of the model are formed by full consensus situations, where all agents share the same opinion, represented by variables $+1$ or $-1$.  The model was after extended by several researchers \cite{Gimenez,pmco,pmco2,krapivsky2021divergence,muslim2021phase,muslim2023}.

In this study we investigate the evolution of agents' opinion in the majority-rule model. We introduce a directed propaganda, oriented directly for the discussion group, that we argue can mimic the complex algorithms and artificial intelligence used in advertisements on social media and online platforms. 
Our results suggest that even a small influence of directed propaganda is enough to avoid consensus in the population. In addition, one can observe an order-disorder phase transition for a specific value of the parameters related to the directed propaganda.

This paper is organized as follows: in Section 2 we present our model as well their dynamic rules. The numerical and analytical results are discussed in Section 3 and our conclusions and final remarks are presented in Section 4. In Appendix one can see the details of the analytical calculations.


\section{\label{sec:model}Model}

\qquad The dynamics of agreement and disagreement is treated in terms of the variation of the number of different opinion states in population, where each individual can have one of two possible opinions ($\pm 1$). The average opinion in the system, represented by $M$, for a population with $N$ individuals, is the order parameter of the system (or the magnetization), defined as

\begin{equation}\label{order_par}
M = \frac{1}{N}\left|\sum_{i=1}^{N} S_i \right| ~,
\end{equation}
in which $S_i$ is the individual opinion for each agent ($i=1,2,...,N$), considered at each time step. We define one time step as an attempt to $N$ agents try to change their opinions by interacting in discussions groups. In our study, for simplicity, we considered a discussion group with a constant size of three agents. Differently from the original model, we are considering an external directed propaganda, acting like an external field, $H = \pm 1$. However, this directed propaganda has no effect on the entire system, but affects only agents on discussion group instead. We argue that the introduction of the directed propaganda as an external field addressed to a specific group of agents can influence the discussion group beyond the majority, leading the entire system to not reach consensus. Furthermore, this introduction of an external directed field can mimic, in a simple way, the actual new forms of online propaganda.

\begin{figure*}[htb]
\centering
\includegraphics[width=0.9\linewidth]{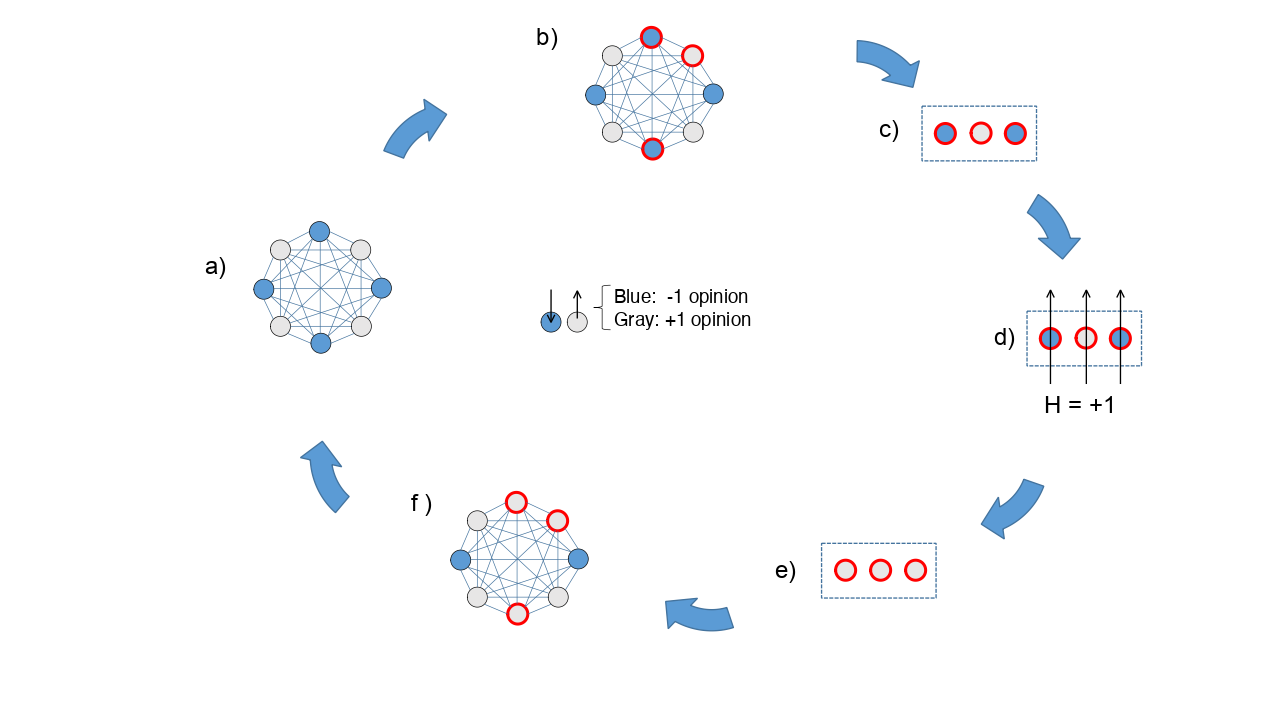}
\caption{\label{fig:model}The schematic representation of our model. From the fully connected population (a) we chose at random (b) the discussion group (c) and, with probability $p$ we set the external field $H = +1$, for instance (d). With probability $q$, all agents follow the external field (e) and the return to the lattice to continue the system dynamics (f).}
\end{figure*}

One can see in Fig. \ref{fig:model} a schematic representation of our model, in which three agents are selected at random from a fully connected graph to form a discussion group. After the action of the external influence or the discussion process, the agents' states are updated and the system carries on with the opinion dynamics. The dynamic rules for our model are very simple and they are explained in the following, as well the updating of individual agents' opinions.

At each time step, we:

\begin{enumerate}
    \item Select at random the three agents to form the discussion group: $S_1$, $S_2$ and $S_3$;
    \item Set the propaganda direction $H=+1$ with probability $p$ and $H=-1$ with probability $1-p$;
    \item With probability $q$ all agents in discussion group follow the propaganda direction $H$, independent of the majority/minority opinion inside the group. So the opinions are updated to $S_1 = S_2 = S_3 = +1$ if $H=+1$ or $S_1 = S_2 = S_3 = -1$ if $H=-1$, depending on the choice of $H$ in the previous step;
    \item On the other hand, with probability $1-q$, the agents follow the majority opinion inside the group: if the opinion of one agent is different from the other two, the agent flips to follow the majority.
\end{enumerate}

\begin{figure}[htb]
\centering
\includegraphics[width=0.8\linewidth]{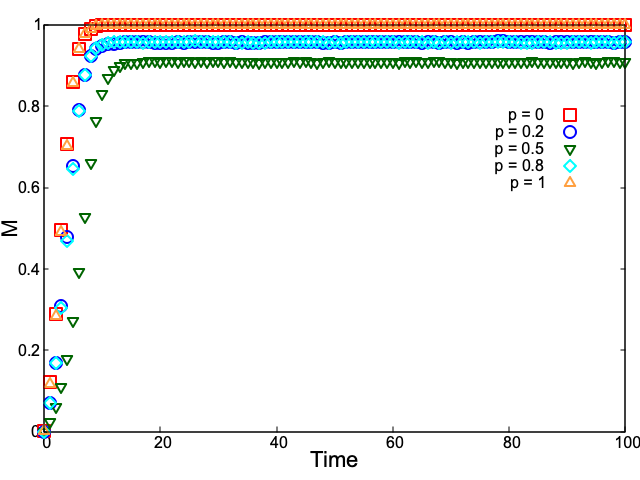}
\caption{\label{fig:Mag_x_time}Average opinion as function of time for $p = 0$, $p = 0.2$, $p = 0.5$, $p = 0.8$ and $p = 1$, for fixed $q=0.1$. Full consensus is reached only for $p = 0$ or $p = 1$.}
\end{figure}

We introduce two parameters, namely $p$ and $q$. The first one, $p$, is the probability of a ``positive'' external propaganda. The parameter $q$ is the probability to the agents in discussion group change their opinions and follow the external propaganda direction. In other words, $p$ is a measure of the propaganda ``volume" for one specific direction and $q$ is a measure of how strong or how effective the propaganda is in order to influence the agents. Our results in the next section will show that, even for a small value of the parameter $q$, the system does not show the full consensus at stationary state that is observed in the original majority-rule model.

\section{Numerical and Analytical Results}

\qquad The introduction of the propaganda in the majority-rule model, acting like an external field and directly influencing the discussion group, changes the stationary state of the model and thus the full consensus can no longer be reached. In fig. \ref{fig:Mag_x_time} we exhibit the time evolution of the average opinion $M$, obtained from numerical simulations of the model, considering Eq. (\ref{order_par}), for $q=0.1$ and typical values of $p$. As one can see in Fig. \ref{fig:Mag_x_time}, for a small value of $q$, the system does not reach a stationary state with full consensus, except for the cases where $~{p = 0}$ or $~{p = 1}$. These situations means that there is no competition between the propaganda, since there is only one possible direction of $H$ and, therefore, it is possible for the system to reach full consensus. Any other case, due to the competition between the two directions of propaganda, a majority opinion (but not totality) is obtained, mimicking a democratic situation. Taking into account an entire population, a very common situation nowadays, in any real situation and for most democratic states, is that there is always a fraction of agents with opinion contrary to the dominant opinion.

\begin{figure*}[htb]
\centering
\includegraphics*[angle=0,width=.33\textwidth]{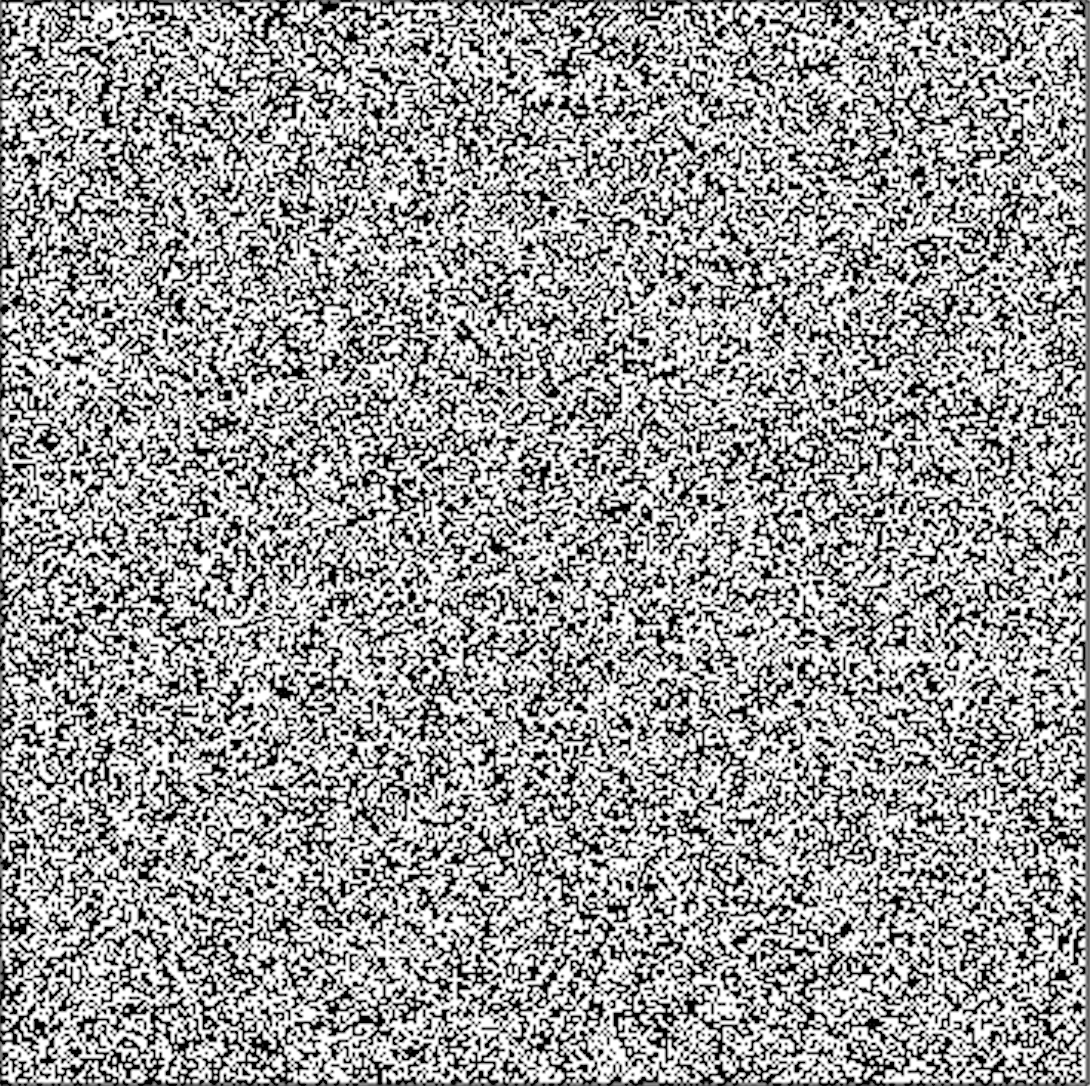}\hfill
\includegraphics*[angle=0,width=.33\textwidth]{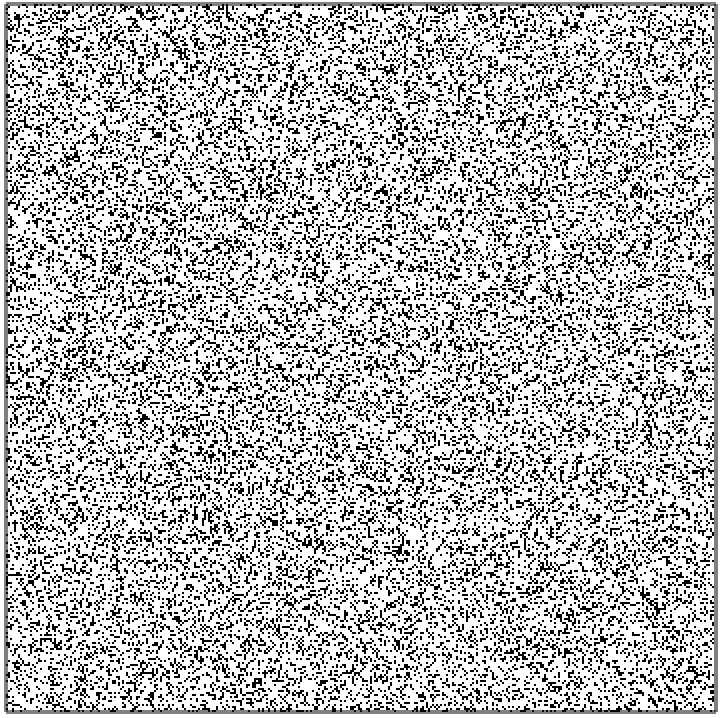}\hfill
\includegraphics*[angle=0,width=.33\textwidth]{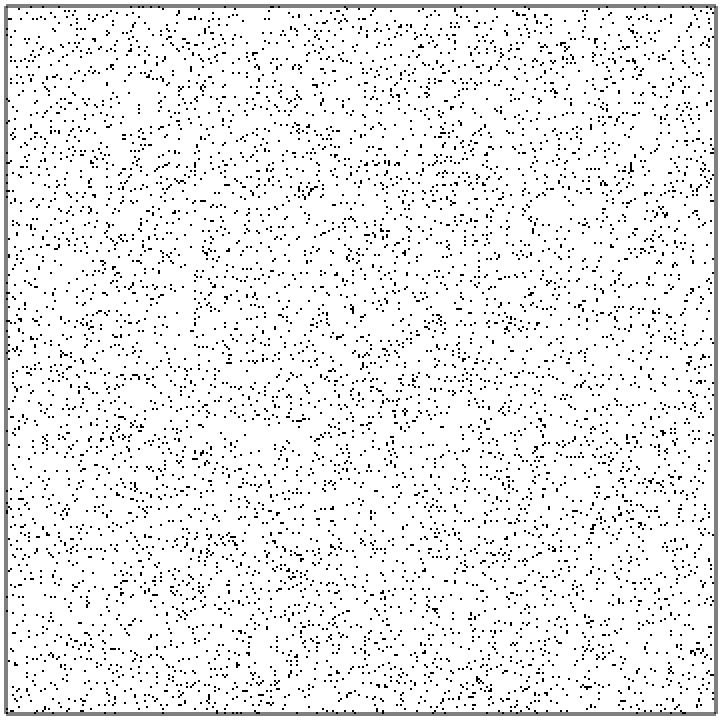}\hfill
\caption{{\label{fig:time_evolution}Evolution of the agents' opinions in our model. Three different instants of time for the same initial configuration is shown. The $-1$ opinion is represented by a black pixel, while the white one represents the $+1$ opinion. We plot a grid $L \times L$ with $L = 500$ agents, and we considered the parameters $p = 0.5$ and $q = 0.3$. As initial configuration, we considered equal fractions for the two opinions, randomly distributed in the grid. From left panel to right, one can see the beginning of the dynamics, an intermediate instant of time and the steady state configuration.}}
\end{figure*}

In our simulations, the system's dynamics initiate with a random distribution of opinions among the agents, i.e., with the same probability each agent can have $+1$ or $-1$ opinion. In Fig. \ref{fig:time_evolution} we show three different instants of time for the agents' opinion. For sufficient long times, the systems reaches the steady state and, even though the individual opinions may change, the average opinion is constant in time.

Since we are considering a fully-connected population, we can consider a mean-field approach to obtain some analytical results for the model. Defining $f_1$ and $f_{-1}$ as the stationary densities of each possible state ($+1$ or $-1$, respectively) and considering the normalization condition where $f_1 + f_{-1} = 1$, analytical expressions can be derived for those densities. By considering the probabilities that a given agent change its opinion ($+1 \rightarrow -1$ or $-1 \rightarrow +1$), we calculate the variations of the average opinion and calculate $M$ at the stationary state (see the Appendix for a detailed description of our calculations).

As discussed in the Appendix, we can obtain a third-order polynomial for the fraction of agents with $+1$ opinion, given by
\begin{equation} \label{eq_2}
2\,(q-1)\,f_{1}^{3} - 3\,(q-1)\,f_{1}^{2} - f_{1} + p\,q = 0~,
\end{equation}
which depends on the two parameters $p$ and $q$. 

\begin{figure}[htb]
\includegraphics*[angle=0,width=.48\textwidth]{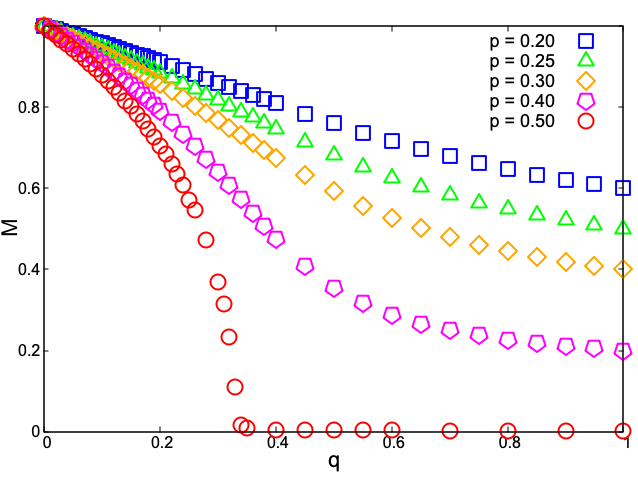}\hfill
\includegraphics*[angle=0,width=.48\textwidth]{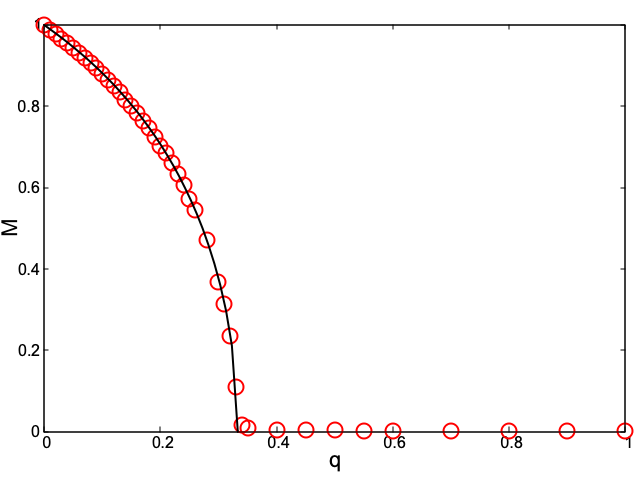}\hfill
\caption{\label{fig:mag_x_q}Steady state average opinion (or the magnetization $M$) versus $q$, for distinct values of p. As one can see on the left panel, when $p = 0.5$, for large enough values of $q$, the system does not present a majority opinion, representing a paramagnetic state where $M=0$. For the other values of $p$ we have $M>0$ for all values of $q$. On the right panel, at the critical point $q_c = 1/3$, an order-disorder nonequilibrium phase transition occurs for the case $p = 0.5$. The solid line is the analytical solution, Eq. \eqref{order_2}, and the red circles are numerical results.}
\end{figure}

For the specific case $p=0.5$, we obtain a first solution given by $f_1 = f_{-1} = 1/2$. The magnetization can be obtained from $M=|f_1 - f_{-1}|$ which gives us $M=0$, representing a paramagnetic state. The other two solutions of the third-order polynomial give us  $f_{1}\neq f_{-1}$ (see the Appendix), which indicates a ferromagnetic phase, where one of the two opinions, $+1$ or $-1$, is the dominant opinion in the population. Considering again the relation $M=|f_{1}-f_{-1}|$ for the mentioned two solutions, we obtain 
\begin{eqnarray} \label{order_2}
M = \frac{\sqrt{3q^{2} - 4q +1}}{q-1} ~,
\end{eqnarray}
which can be rewritten in the critical phenomena language as $M \sim (q_c-q)^{\beta}$, where the critical point is $q_c = 1/3$ and the order parameter critical exponent is $\beta=1/2$, as shown in Appendix. The order-disorder nonequilibrium phase transition at $q_c$ seems to be in the Ising mean-field universality class, as it is usual in opinion models with two or three states \cite{bcs_review}. Our model only presents a phase transition for the case $p = 0.5$, as it is discussed in the Appendix.

In fig. \ref{fig:mag_x_q} we exhibit the numerical results for the steady-state magnetization for different values of $q$. The critical point $q_c$ obtained analytically for $p=0.5$ is in agreement with our numerical results, as one can see in the right panel of fig. \ref{fig:mag_x_q}. For the other values of $p$, we observe no phase transition, as discussed in the Appendix. For such case, we observe that one of the two opinions, $+1$ or $-1$, is the majority opinion in the population, for all values of $p$ and $q$. In such a case, we observe that increasing $q$, i.e., increasing the effect of the directed propaganda in the population, leads to more polarized states (lower values of $M$).

\begin{figure}[htb]
\includegraphics[width=\linewidth]{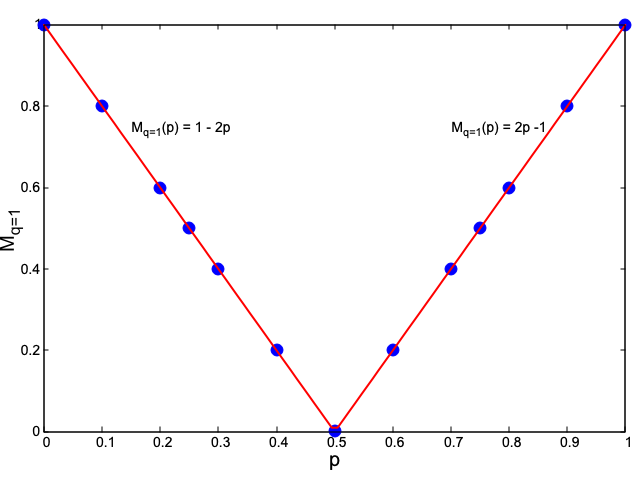}
\caption{\label{fig:mag_q=1x_p}Steady state average opinion (or the magnetization $M$) versus $p$ for the case $q=1$. The blue circles were obtained from the numerical simulations of the model, whereas the solid red line is the analytical result $M=|2p-1|$. Except for $p=0.5$, one opinion is always the dominant opinion in the population. Full consensus are observed for $p=0$ and $p=1$.}
\end{figure}

We also studied the steady-state magnetization for the case $q = 1$ and different values of $p$. Considering $q=1$ in Eq. (\ref{eq_2}), one can obtain $f_1 = p$, which by the normalization condition leads to $f_{-1}=1-p$. Thus, the stationary average opinion is $M=|f_1-f_{-1}|=|2p-1|$. Our numerical simulations are in agreement with the analytical calculation. As one can see in fig. \ref{fig:mag_q=1x_p}, except for the case when $p = 0.5$, one of the two possible opinions is the dominant position in the population. Exceptions are observed for the extreme cases $p=0$ and $p=1$. Considering the analytical results, we see that for $p=0$ the stationary state of the system exhibits a population with all agents sharing opinion $-1$ ($f_{-1}=1$), whereas for $p=1$ the final state of the population is a full consensus with all agents sharing the same opinion $+1$ ($f_{1}=1$). These consensus situations are easily to be understood. Indeed, for $q=1$ the majority rule will not be applied, and the agents in the discussion group will always follow the media effect. For the case $p=0$, the directed propaganda will always favors opinion $-1$, leading all agents to follow the media opinion after a long time. Analogously, for $p=1$ the directed propaganda will always favors opinion $+1$, leading to another situation of full consensus at the steady states.

Summarizing, we observe full consensus states ($M=1$) only in special situations ($q=0$ for any value of $p$, and $q=1$ for $p=0$ and $p=1$). Usually in extensions of the original majority-rule model \cite{GALAM1999132} the goal is to study such consensus situations and how the population reaches the consensus states \cite{PhysRevLett.90.238701,pmco,muslim2021phase,muslim2023}. Here, we included a directed propaganda in a simple way, and the results show that we can observe the phenomenon of polarization of opinions considering a simple mechanism.

\section{\label{sec:fin-rem}Final remarks}

\qquad We studied the majority-rule model with the inclusion of an external propaganda (external field) acting directly to the discussion group instead the entire population. The local group propaganda can favors opinion $+1$ with probability $p$ and opinion $-1$ with probability $1-p$. After that, all agents in the discussion group follow the media opinion with probability $q$, or with the complementary probability $1-q$ the agents follow the standard majority rule. We argue that the directed propaganda can mimic in a very simple way the complex algorithms and artificial intelligence used in advertisements on social media and online platforms. The inclusion of the directed propaganda modify the original majority-rule model, introducing an order-disorder phase transition. We show through analytical and numerical results that this transition occurs at $q_c = 1/3$ for the case $p = 0.5$. For $p\neq 0.5$ this transition is absent. Our results show that the usual majority-rule model stationary state is reached, with a full consensus, for the case when the external propaganda is absent ($q=0$). In the other limiting case $q=1$, where the agents will always follow the directed propaganda, the agents can reach full consensus for the extreme cases $p=0$ and $p=1$. However, even for a small influence of the external propaganda, the final state is reached with majority but not full consensus. For the case which the propaganda influence is strong enough among the agents, we show that the consensus can not be reached at all. In other words, the consideration of directed propaganda, in a simple way, in the majority-rule model favors the polarization of opinions, and in the symmetric case $p=0.5$ we observe full polarization (magnetization $M=0$).

\appendix
\section{Analytical calculations}

Following the approach of References \cite{pmco,biswas_2012,nuno_celia_2012}, we computed the stationary order parameter $M$. Let us first define $f_{1}$ and $f_{-1}$ as the stationary probabilities of each possible state ($+1$ or $-1$, respectively), in a way that we have the normalization condition
\begin{equation}\label{normal}
f_{1} + f_{-1} = 1 ~.
\end{equation}
We have to calculate the probability that a given agent suffers the change $+1\to -1$ or $-1\to +1$. We are considering groups of 3 agents, so one can have distinct variations of the magnetization, depending on the states of the 3 agents and on the probabilities $p$ and $q$. For example, the probability to choose at random 3 agents with opinions $o=+1$, i.e, a configuration $(+,+,+)$, is $f_{1}^{3}$. With probability $q$ the agents will follow the local media effect. In addition, with probability $p$ the configuration remains  $(+,+,+)$, which does not affect the magnetization of the system, since the local magnetic field in this case is given by $H=+1$. However, with probability $1-p$ the local field will be given by $H=-1$, so the group will change opinion to $(-,-,-)$, which cause a variation of $-6$ in the magnetization. In other words, the magnetization decreases $6$ units with probability $q\,(1-p)\,f_{1}^{3}$. One can denote this probability as $r(-6)$, i.e., the probability that the magnetization variation is equal to $-6$. Generalizing, one can define $r(k)$, with $-6\le k \le +6$ in this case, as the probability that the magnetization variation is $k$ after the application of the model's rules. As the order parameter (magnetization) stabilizes in the steady states, we have that its average variation must vanish in those steady states, namely,
\begin{equation} \label{nullshift1}
6\,[r(+6)-r(-6)] + 4\,[r(+4)-r(-4)] + 2\,[r(+2)-r(-2)]=0 \,.
\end{equation}

\noindent
In this case, we have
\begin{eqnarray} \nonumber
r(+6) &=& p\,q\,f_{-1}^{3} \\ \nonumber
r(-6) &=& (1-p)\,q\,f_{1}^{3} \\ \nonumber
r(+4) &=& 3\,p\,q\,f_{1}\,f_{-1}^{2} \\ \nonumber
r(-4) &=& 3\,(1-p)\,q\,f_{1}{2}\,f_{-1}^ \\ \nonumber
r(+2) &=& 3\,p\,q\,f_{1}^{2}\,f_{-1} + 3\,(1-q)\,f_{1}^{2}\,f_{-1} \\ \nonumber
r(-2) &=& 3\,(1-p)\,q\,f_{1}\,f_{-1}^{2} + 3\,(1-q)\,f_{1}\,f_{-1}^{2} ~.
\end{eqnarray}
Thus, the null average variation condition Eq. (\ref{nullshift1}), together with the normalization condition Eq. (\ref{normal}) written in the form $f_{-1} = 1 - f_{1}$, gives us a third-order polynomial for $f_{1}$, namely
\begin{equation} \label{eq_f1}
2\,(q-1)\,f_{1}^{3} - 3\,(q-1)\,f_{1}^{2} - f_{1} + p\,q = 0 ~.
\end{equation}
The polynomial gives us three distinct solutions. 

Let us first consider the specific case $p=0.5$. The mentioned three solutions are given by 
\begin{eqnarray}\label{eqf1_1}
f_{1} &=& \frac{1}{2} \\ \label{eqf1_2}
f_{1} &=& \frac{(q-1)\pm\sqrt{3q^{2} - 4q +1}}{2(q-1)} ~.
\end{eqnarray}
The first solution, $f_{1}=1/2$, leads to $f_{-1}=1/2$ due to the normalization condition, Eq. (\ref{normal}). Thus, $f_{1}=f_{-1}$ indicates a paramagnetic phase. The other 2 solutions lead to $f_{1}\neq f_{-1}$, which indicates a ferromagnetic phase, where one of the two opinions, $+1$ or $-1$, is the majority opinion in the population. As Eq. (\ref{eqf1_2}) predicts two solutions (see the $\pm$ signals), one has two curves for each value of $q$ for $q<q_c$, where $q_c$ is the critical point, that will be obtained in the following. When $f_{1}$ assumes one of the values in Eq. (\ref{eqf1_2}), consequently $f_{-1}$ takes the other one \cite{nuno_allan}.

The order parameter can be obtained from $M=|f_{1}-f_{-1}|$, which give us
\begin{eqnarray}\label{order}
M = \frac{\sqrt{3q^{2} - 4q +1}}{q-1} = \frac{\sqrt{3(1-q)(1/3-q)}}{q-1} \sim (q_c-q)^{\beta} ~,
\end{eqnarray}
where $q_c(p=1/2) = 1/3$ and $\beta=1/2$. Eq. (\ref{order}) is valid for $q<q_c$ and it shows an order-disorder nonequilibrium phase transition at $q_c=1/3$, and the mentioned transition appears to be in the Ising mean-field universality class, as is usual in opinion models with two or three states \cite{bcs_review}.

For $p\neq 0.5$, the general solution of Eq. (\ref{eq_f1}) leads to 2 complex conjugate solutions, and the third solution is a real function for $f_1$. In such a case, the absence of a solution $f_{1}=1/2$ indicates the absence of a paramagnetic phase, and thus there is no phase transition since we always will have $f_1 \neq f_{-1}$, leading to a ferromagnetic state where one of the two opinions is the majority opinion in the population ($M>0$). Thus, the model only presents a phase transition for the case $p=1/2$.

\section*{Acknowledgments}
NC acknowledges  financial  support  from  the  Brazilian  funding  agencies  Conselho Nacional de Desenvolvimento Cient\'ifico e Tecnol\'ogico (CNPq, Grant 310893/2020-8) and Funda\c{c}\~ao Carlos Chagas Filho de Amparo \`a Pesquisa do Estado do Rio de Janeiro (FAPERJ, Grant 203.217/2017).

\bibliographystyle{spphys}       
\bibliography{main.bib}

\end{document}